\documentclass{nature}
\pdfoutput=1
\usepackage{graphicx,subfig}
 \usepackage[normalem]{ulem}
\usepackage{float}
\usepackage[usenames]{color}
\usepackage{amsmath}
\usepackage{booktabs}
 \usepackage{multirow}
\usepackage{blindtext}
\usepackage{longtable}
\usepackage{textcomp}
\usepackage{color}
\usepackage{array}
\usepackage{url}
\usepackage[hidelinks]{hyperref}
\hypersetup{
    colorlinks,
    linkcolor={blue},
    citecolor={blue},
    urlcolor={blue}
}

\begin{document}

\title{Experimental evidence of symmetry breaking of transition-path times}
\author{J. Gladrow$^{1}$, M. Ribezzi-Crivellari$^{2,3}$, F. Ritort$^{3,4}$, \& U. F. Keyser$^{1*}$}

\maketitle
\begin{affiliations}
 \item Cavendish Laboratory, University of Cambridge, Cambridge CB3 0HE, United Kingdom
 \item Laboratoire de Biochimie (LBC), ESPCI Paris, PSL Research University, CNRS UMR8231 Chimie Biologie Innovation, Paris, France
 \item Condensed Matter Physics Department, University of Barcelona, C/Marti i Franques s/n, 08028 Barcelona, Spain
 \item CIBER-BBN de Bioingenieria, Biomateriales y Nanomedicina, Instituto de Salud Carlos III, 28029 Madrid, Spain
 \item[*] Corresponding author: Ulrich F. Keyser, ufk20@cam.ac.uk
 \end{affiliations}

\section*{Abstract}
\vspace{0.5cm}
\begin{abstract}

While thermal rates of state transitions in classical systems have been studied for almost a century, associated transition-path times have only recently received attention.
Uphill and downhill transition paths between states at different free energies should be statistically indistinguishable. Here, we systematically investigate transition-path-time symmetry and report evidence of its breakdown on the molecular- and meso-scale out of equilibrium. In automated Brownian dynamics experiments, we establish first-passage-time symmetries of colloids driven by femtoNewton forces in holographically-created optical landscapes confined within microchannels. Conversely, we show that transitions which couple in a path-dependent manner to fluctuating forces exhibit asymmetry. We reproduce this asymmetry in folding transitions of DNA-hairpins driven out of equilibrium and suggest a topological mechanism of symmetry breakdown. Our results are relevant to measurements that capture a single coordinate in a multidimensional free energy landscape, as encountered in electrophysiology and single-molecule fluorescence experiments. 

\end{abstract}
\maketitle

 %%  Thermodynamic states are conceptualized as areas in phase-space that are located around certain minima of the equilibrium free energy $U$. Importantly, state areas do not necessarily adjoin each other, such that each transition is associated with a time of travel. 
\noindent 
\section*{Introduction}
%%% FIRST PART: Background, context and rationale of the study.
Classical thermally activated reactions are ubiquitous in nature and technology with a wide range of examples including folding transitions of proteins\cite{Schuler2002, Chung2009,Chung2012,Chung2015} and DNA\cite{Woodside2006,Neupane2016}, transitions of colloidal particles between optical traps\cite{McCann1999}, and the dynamics of molecules in membrane channel proteins\cite{Nestorovich2002,Pages2008,Stein2014}, artificial nanopores\cite{Dekker2007,Bell2012}, and channels\cite{Yang2017a, Skaug2018}. Depending on the system, these transitions may proceed along a single, multiple, or even a continuum of pathways in phase-space that connect the initial and final states. The question of pathway multiplicity is, for instance, currently debated in the context of protein folding\cite{Englander2017a, Eaton2017, Englander2017}. The transition-path time $\tau$ is defined as the time it takes to travel from one thermodynamic state to another. $\tau$ and especially its distribution contains valuable information about the pathway and the underlying system.
In a two-state system, the transition-path time $\tau_{\rm{I} \rightarrow \rm{II}}$ can be measured as follows: whenever the system leaves the area of state I, a stopwatch is triggered. It is stopped when the system either returns or transitions to state II. In the latter case, this constitutes a single realization of the transition path.
Importantly, transition-path times $\tau$ do not directly determine the rates $k_{\rm{I} \to \rm{II}}$ and $k_{\rm{I} \leftarrow \rm{II}}$ of the reaction\cite{Kramers1940, Hanggi1990}. This is because, rate coefficients account for all prior unsuccessful attempts at leaving the state in addition to the actual time of travel of the successful attempt. Transition rates can therefore be strongly asymmetric if one state is thermodynamically favourable over the other. By contrast, transition-path times are expected to be statistically symmetric in equilibrium, in accord with the principle of microscopic reversibility\cite{Onsager1953, Berezhkovskii2006, ALEXKAMENEV2011}.

In many systems, it has been challenging to resolve individual transitions. However, technological progress has now advanced to a point where information about folding events of polymers can be gathered using optical techniques, such as F\"{o}rster resonance energy transfer (FRET), which has sparked considerable interest in transition-path times\cite{Chung2009,Chung2012,Chung2015}. Time-resolved force spectroscopy based on optical tweezers has been successfully applied in studies of the folding pathways of proteins and DNA\cite{Borgia2008, Gupta2011, Yu2012, Neupane2016, Neupane2017, Cossio2018}. Measurements of ribosomal stepping times along RNA have led to insights into the molecular mechanics of gene translation\cite{Wen2008}. The same technique has also been used to show a transition-path time symmetry in equilibrium folding and unfolding transitions of DNA-hairpins\cite{Neupane2016}.
Recently, local velocities along folding trajectories of DNA have been measured with high resolution of the folding coordinate, which shed light on the frequency of recrossing events in relation to all barrier crossings\cite{Neupane2018}.

In electrophysiology and pulse-sensing experiments, molecules are interrogated by voltage-driven transport often proceeding along a single pathway through membrane channels\cite{Kasianowicz1996}. This technique detects changes in ion flow due to blockage by solutes of interest. However, for small, uncharged molecules (such as some antibiotics), these measurements are often not sensitive to the direction of travel or the orientation of the channel\cite{Mahendran2009}. Direction, however, matters in biological membrane channels, which usually have an asymmetric structure. Channel asymmetry can, for instance, give rise to ratcheting effects that rectify diffusive currents under the influence of fluctuating forces\cite{Kosztin2004}. Despite their importance, most of the thermodynamic principles of transition-path times have not been studied systematically, especially in experiments outside of thermodynamic equilibrium.
%Interestingly, the theoretical framework for transition path times\cite{Daldrop2016} in molecular folding\cite{Truex2015, Neupane2016a} also describes translocation times of solutes that are interacting with membrane channels\cite{Hilty2001, Berezhkovskii2003}.

\begin{figure}
 \includegraphics[width=18cm]{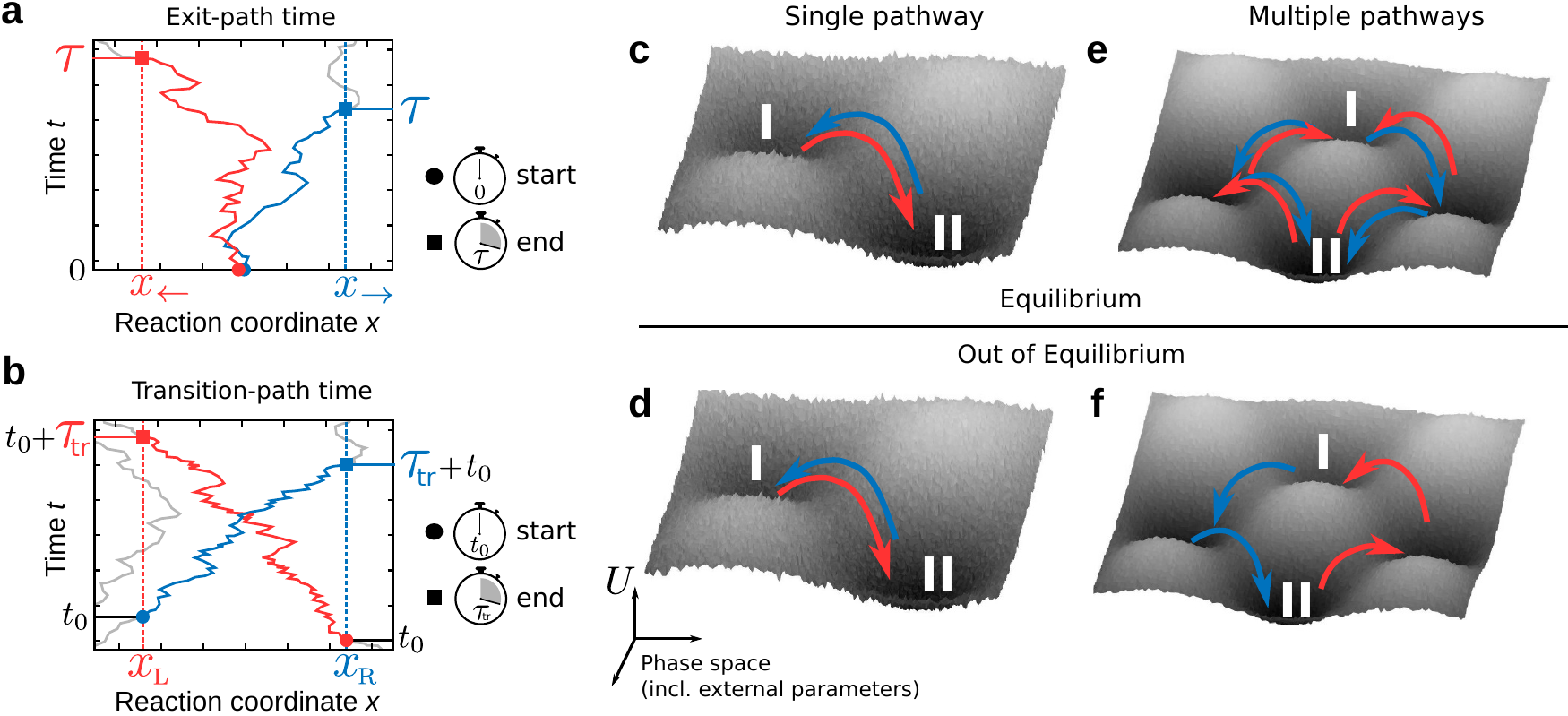}
  \caption{\label{fig:introSketch} Illustrations of conditional exit- and direct transition-path times and of pathway multiplicity with and without detailed balance. {\bf a} The exit-path time $\tau$ to a given exit from a spatial interval is defined as the time of first-passage of that boundary (dotted line) before other boundaries are reached after initialization within the interval. {\bf b} Transition-path times $\tau_\mathrm{tr}$ between the boundaries of an interval are first-passage times of trajectories that start from one boundary and directly reach the opposite boundary. Trajectories that return to the same boundary (see grey trajectory) are not counted. As we explain in the text, we deliberately named the transition-path boundaries $x_\mathrm{L}$ and $x_\mathrm{R}$ to distinguish them from exit-path boundaries. {\bf c} Illustration of (mean) transition pathways (arrows) in potential landscapes with two reaction coordinates and a single pathway. {\bf d} Sketch of out of equilibrium transitions that follow the same pathways and consequently do not break detailed balance. {\bf e} Sketch of transitions that are time symmetric in equilibrium, despite pathway multiplicity. {\bf f} The transition-path time symmetry may break down if transition paths diversify out of equilibrium.}
\end{figure}

%  Furthermore, we show the equivalence of transition path time distributions in time-constant potential landscapes and demonstrate their consistency with predictions that assume absorbing boundary conditions located at the initial and end points of the transition (see Fig.\ref{fig:introSketch}{\bf b}).
%%%% SECOND PART: SINGLE PARAGRAPH, RESULTS AND FINDINGS OF THE STUDY. MORE DETAILED INFORMATION ON THE APPROACH OR THE PARTICULARS OF THE STUDY SHOULD APPEAR IN 
%%%% THE OPENING SECTION OF RESULTS OR EVEN METHODS, as appropriate.
We begin by showing experimentally that a pathway symmetry in steady-state potentials is reflected in a robust and measurable symmetry in path times. Counterintuitively, a symmetry $\langle \tau_{\rm{I} \to \rm{II}} \rangle=\langle \tau_{\rm{I} \leftarrow \rm{II}}\rangle$ would imply steeper potential gradients between states I and II to not only expedite downhill, but also uphill transitions. We demonstrate two flavours of this symmetry: (i) for exit paths from a one-dimensional spatial interval that are conditioned on a particular exit\cite{Stern1977, Neri2016}(see Fig.\ref{fig:introSketch}{\bf a}) and (ii) direct transition paths across an interval (see Fig.~\ref{fig:introSketch}{\bf b}). Our fully-automated setup consists of a holographic optical tweezers (HOT) setup, which creates potential landscapes for colloidal particles within microfluidic channels. Conceptually, our results show that transition-path times are well-defined theoretically when measured from uninterrupted trajectories between any two points in phase-space, regardless of whether or not these points lie in thermodynamic states defined by deep potential minima. In the second part of the study, we show that a breakdown of detailed balance in a coarse-grained, multidimensional phase space can lead to a breakdown of transition time symmetry (see Fig.~\ref{fig:introSketch}{\bf f}). We demonstrate the breakdown of this symmetry in microscopic transitions in asymmetric, bistable potentials, perturbed by fluctuating forces. Then, we show that this breakdown of symmetry extends to the molecular scale. Specifically, we explore the kinetics of folding and unfolding transitions of a DNA-hairpin grafted onto colloidal particles driven out of equilibrium by telegraphic noise. To the best of our knowledge, properties of transition times under time-dependent forces have not been described experimentally so far.

\section*{Results}
\label{sec:results}

\subsection{Uphill and downhill exit-path-time symmetry} 
\begin{figure}
 \includegraphics[width=\textwidth]{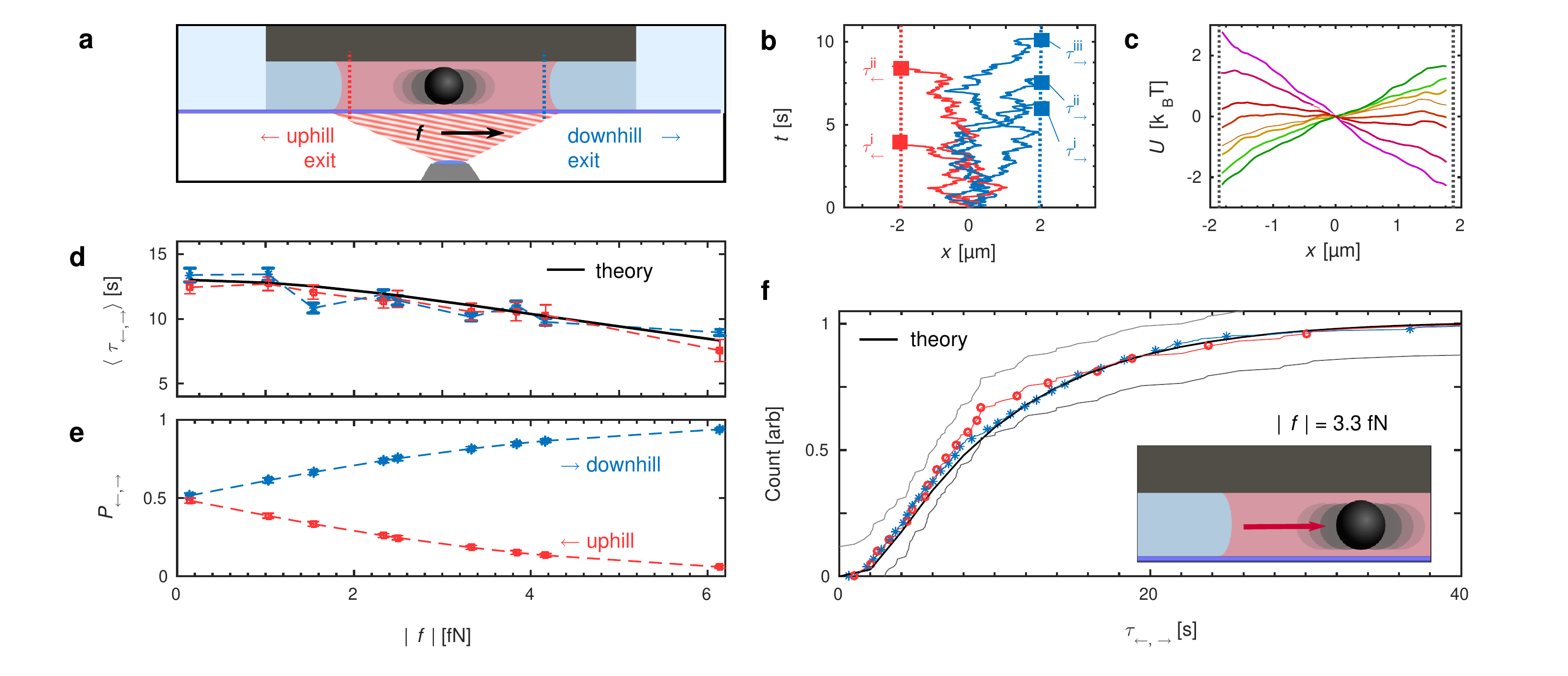}
  \caption{\label{fig:microfluidicPanel} Exit-path times in the uphill and downhill direction are identically distributed. {\bf a} Exit-path times of a colloidal particle from a predefined interval (dotted lines) are measured for various phase-gradient forces $f\approx 0-6$ fN. {\bf b} Examples of five trajectories and exit-path times. {\bf c} Inferred potentials (see Methods). The standard error of the mean is given by the line width. {\bf d} Mean exit-path times in the uphill (red) and downhill (blue) directions agree, but decrease with increasing force. Error bars indicate the standard error of the mean. {\bf e} Probability to exit the interval in the uphill (red) and downhill (blue) direction. Error bars (on the order marker size) show the standard error of the mean. {\bf f}  Cumulative distribution of first exit-path times in the uphill (red) and downhill (blue) direction for the $3.3$ fN-data point and boundaries of the Kolmogorov-Smirnov test (light grey). The exit times were measured from a predefined interval similar to panel {\bf a} (dotted lines). While the lines represent the full data set, the markers only show a selection of the data to avoid crowding of the figure. The black line in {\bf d} and {\bf f} show Fokker-Planck-based predictions as discussed in the text.}
\end{figure}

We use our HOT setup in conjunction with confining microchannels to physically simulate the escape of a Brownian particle from a cavity, reminiscent of the escape of solutes from membrane channels\cite{Pagliara2013}. The movements of solutes such as ions in membrane channels or nanopores often follow thermodynamic gradients. Such a gradient is modelled here with a phase-gradient force\cite{Curtis2012} $f$ in a laser line trap as sketched in Fig.~\ref{fig:microfluidicPanel}{\bf a}. We plot a selection of trajectories of Brownian particles acquired from automated drag-and-drop experiments in panel \ref{fig:microfluidicPanel}{\bf b}: At $t=0$, the particle is positioned at the centre $x_0=0$ of a predefined spatial interval within a microchannel. The particle is released, a stopwatch is triggered, and a laser line trap with a prescribed phase-gradient is turned on (see Fig.~\ref{fig:introSketch}{\bf a} and Methods). Once the particle leaves the interval, the stopwatch and the measurement are stopped. We find that the probability density of positions $\rho(x)$ recorded in an ensemble of repeats goes to zero at the interval boundaries $x_\leftarrow$, $x_\rightarrow$. These points can therefore be considered as absorbing in the Fokker-Planck picture. For each value of force $f$, we gathered around $1000$ trajectories of which $950-980$ were free of incidents such as other particles entering the channel. We applied both positive and negative forces (see Fig.~\ref{fig:microfluidicPanel}{\bf c}) to check for static bias caused for example by weak latent flows of water. Within our experimental resolution, we find that the diffusion profile $D(x)$ along the channel is not affected by the applied phase-gradients (see Supplementary Note 2 and Supplementary Figure 2), such that any difference in dynamics must be attributed to the difference in force. The inference method used to estimate potential $U(x)$ and diffusion profile $D(x)$ is discussed in the Methods section.

The central result of this experiment is the equivalence of the mean left and right exit-path times $\langle \tau_\leftarrow \rangle$, $\langle \tau_\rightarrow \rangle$, shown in Fig.~\ref{fig:microfluidicPanel}{\bf d}. Moreover, shorter exit-path times for higher absolute forces $|f|$ indicate a speed-up of both uphill and downhill trajectories. By contrast, exit probabilities behave intuitively: exits against the force (uphill direction) become increasingly unlikely with an increasing force magnitude $|f|$ as shown in Fig.~\ref{fig:microfluidicPanel}{\bf e}.

The theoretical mean exit-path time (black line in Fig.~\ref{fig:microfluidicPanel}{\bf d}) was obtained from a solution of the mean first-passage time equation $ D\langle\tau\rangle ''(x)+\frac{f}{\gamma} \langle\tau\rangle'(x) = -1$ with boundaries at $x_\leftarrow=-L/2$ and $x_\rightarrow=L/2$ with $ L= 3.7$ \textmu{}m. We note that $L$ stands for the length of the interval, not the channel (see Methods). Due to the observed exit time symmetry, the equation for the mean exit time can be solved for any exit $x_\leftarrow$, $x_\rightarrow$. Interestingly, this symmetry extends beyond a simple equivalence of the mean. In fact, the distribution of exit-path times agree in the uphill and downhill direction as shown in Fig.~\ref{fig:microfluidicPanel}{\bf f}. This holds even for forces of different nature such as hydrodynamic drag (see Supplementary Note 3 and Supplementary Figure 3). Trajectories that manage to exit against the flow do so at precisely the same drift speed $\langle \dot{x}\rangle$ as the ones that follow the flow. The statistical significance of similarity between two given cumulative distributions is asserted by the Kolmogorov-Smirnov test. Throughout this study we require a significance of $0.5$.
Theoretical distributions of exit-path times shown in Fig.~\ref{fig:microfluidicPanel}{\bf f} and Supplementary Figure 3{\bf a} to, for example, the left side $\rho_{\tau_\leftarrow}(t)$ were obtained from a numerical solution of the Fokker-Planck equation $\partial_t \rho(x,t |x_0)= -\partial_x j(x,t|x_0)$ with $j(x,t|x_0) = f\rho(x,t|x_0)-D(x) \partial_x \rho(x,t|x_0)$ denoting the current of probability. 
The initial density of colloid positions $\rho(x,t_0=0) =\rho_0(x)$ was modelled as a sharp peak at the channel centre, $x_0=0$. Once $j(x,t |x_0)$ is obtained, the exit-path time distribution is given by  $\rho_{\tau_\leftarrow}(t)=j(x_\leftarrow, t |x_0)/P_\leftarrow(x_0)$, with $x_\leftarrow$ denoting the $x$-position at the left boundary\cite{Gardiner2009}. Both boundaries were treated as absorbing, i.e. $\rho(x_\leftarrow,t)=\rho(x_\rightarrow,t)=0$. The two exit probabilities $P_\leftarrow(f, x_0=0)$ and $P_\rightarrow(f, x_0=0)$ read $P_\rightarrow(f)=(1+\exp(\frac{fL}{2k_B T}))^{-1}$ and $P_\leftarrow(f)=1-P_\rightarrow(f)$\cite{Gardiner2009}. $\rho_{\tau_\rightarrow}(t)$ can be obtained by exchanging $x_\leftarrow$ for $x_\rightarrow$ and $P_\leftarrow$ for $P_\rightarrow$.

%Interestingly, a time-reversal of a downhill-exiting trajectory will yield an uphill trajectory, which may, however, exit the interval at a different time. For instance, the time-reversed version of the trajectory associated with $\tau_\rightarrow^{iii}$ (see Fig.\ref{fig:microfluidicPanel}{\bf B}) would exit on the uphill side, but earlier. This can be remedied by resetting the stopwatch whenever the trajectory crosses $x=0$. For symmetric channels, this would effectively turn exit path times into transition-path times.

\subsection{Uphill and downhill transition-path-time symmetry}
\begin{figure} 
 \centering
 \includegraphics[width=9cm]{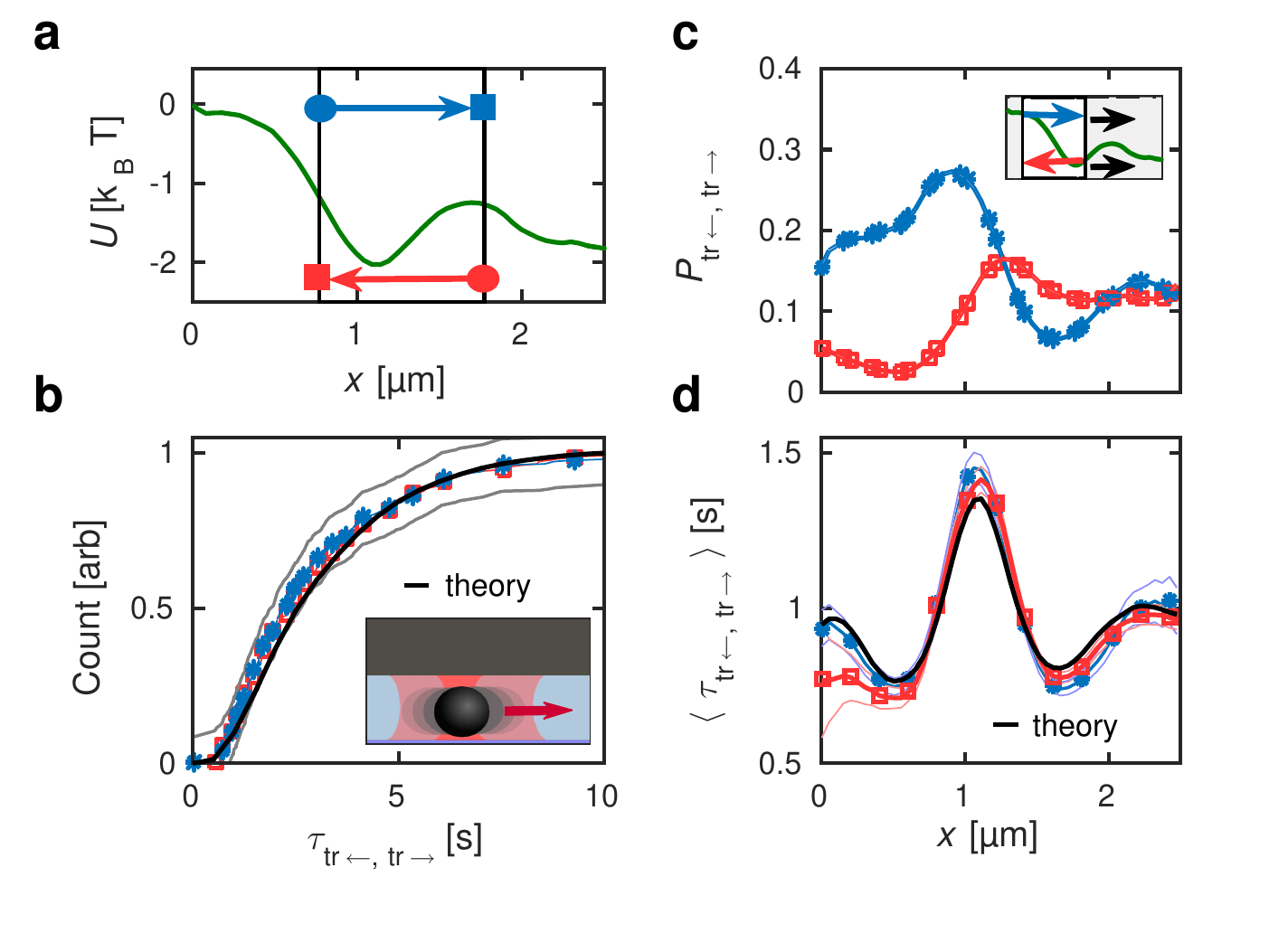}
  \caption{\label{fig:directTransitionTimes} Transition-path-time symmetry of particles in force landscapes. {\bf a} Free energy $U(x)$ calculated from inferred forces (see Methods). The standard error of the mean (not shown) is smaller than the line thickness. {\bf b} Cumulative probability of transition-path times across the region defined by the black box in {\bf a} (compare with Fig.\ref{fig:introSketch}B). Kolmogorov-Smirnov test boundaries are plotted in light grey. {\bf c} Transition probabilities across a $1$ \textmu{}m region, which is continuously moved along $x$. The standard error of the mean (not plotted) is on the order of the marker size. {\bf d} Transition path path times over the same spatial interval. Error envelopes show the standard error of the mean. While the lines in panels {\bf b} - {\bf d} show the full data set, the markers only highlight selected data points.}
\end{figure}

The robust symmetry observed in exit-path times is also found in direct transitions between any two points $x_\mathrm{L}$, $x_\mathrm{R}$ in a quasi-one-dimensional microchannel that is filled with an optical landscape. We deliberately choose left ($\mathrm{L}$) and right ($\mathrm{R}$) subscripts here to contrast transition-path times from exit-path times (see also Fig.~\ref{fig:introSketch}{\bf a},{\bf b}). The landscape considered here consists of a mixture of a point trap and a line trap with a positive phase-gradient force created by our HOT (see inset in panel \ref{fig:directTransitionTimes}{\bf b}). We used the same HOT automation routine as before to observe around 500 uninterrupted colloid trajectories. The energy potential inferred from this ensemble of particle trajectories is plotted in panel \ref{fig:directTransitionTimes}{\bf a}. The transition-path times $\tau_{\text{tr}_\leftarrow}$, $\tau_{\text{tr}_\rightarrow}$ across the interval shown as a black box in \ref{fig:directTransitionTimes}{\bf a}, are identically distributed as can be seen in \ref{fig:directTransitionTimes}{\bf b}. Based on a spline interpolation of the inferred potential $U(x)$ and a spatially dependent diffusion coefficient $D(x)$, we calculated the theoretical distribution of transition-path times $\rho_{\tau_\text{tr}}(t)$. Again, we treat both boundaries  $x_\mathrm{L}$ and $x_\mathrm{R}$  as absorbing. Following Zhang {\em et al.}\cite{Zhang2013}, we compute $\rho_{\tau_\text{tr}}(t)$ for an initial density $\rho_0(x)$ which is sharply peaked close to the initial exit. For the sake of this example, we choose the direction left to right and thus set $x_0(\epsilon)=x_\mathrm{L}+\epsilon$. The current density reads $j(x_\mathrm{R},t|x_0)=\partial_x (U(x) \rho(x,t|x_0))-D(x_\mathrm{R})\partial_x \rho(x,t|x_0)$ at the right boundary $x_\mathrm{R}$. We normalize the distribution by the overall probability to exit through $x_\mathrm{R}$, $P_\rightarrow(x_0)=\int_{x_\mathrm{L}}^{x_0}\, \mathrm{d}x  e^{U(x)/k_B T}/\int_{x_\mathrm{L}}^{x_\mathrm{R}}\, \mathrm{d}x  e^{U(x)/k_B T}$, assuming $x_0$ as the initial position. Finally, we obtain the distribution of transition times $\rho_{\tau_\text{tr}\rightarrow}(t)$ from
\begin{align}
 \rho_{\tau_\text{tr}\rightarrow}(t) = \lim\limits_{\epsilon \searrow 0} \frac{j\left(x_\mathrm{R}| x_0(\epsilon) \right)}{P_\rightarrow\left(x_0(\epsilon)\right)}. \label{eq:transitionTimeDistribution}
\end{align}
A plot of $\rho_{\tau_\text{tr}\rightarrow}(t)$ is shown in Fig.~\ref{fig:directTransitionTimes}{\bf b} (black).

In panel \ref{fig:directTransitionTimes}{\bf c} we plot the probability of direct transition across the same interval length, when this interval is continuously moved along the channel. For each position of this interval, the transition probabilities and times are recorded. As can be seen in panel \ref{fig:directTransitionTimes}{\bf d}, the mean transition-path times calculated in this way in both directions are sensitive to the local force, especially when the transition interval touches the optical point trap. Despite this sensitivity, transition times in both directions are in excellent agreement. 
The theoretical prediction for the mean transition-path time $\langle \tau_{\mathrm{tr}_\rightarrow}\rangle$ plotted in black in Fig.~\ref{fig:directTransitionTimes}{\bf d} was calculated using\cite{Berezhkovskii2017} 
\begin{align}
 \langle \tau_{\mathrm{tr}_\rightarrow}\rangle = \frac{\int_{x_\mathrm{L}}^{x_R}\, \mathrm{d} x \left(\int_{x_\mathrm{L}}^{x}\, \mathrm{d}x' e^{\frac{U(x')}{k_B T}}/D(x')\right) \left(\int_{x}^{x_\mathrm{R}}\, \mathrm{d}x' e^{\frac{U(x')}{k_B T}}/D(x') \right) e^{-\frac{U(x)}{k_B T}}}{ \int_{x_\mathrm{L}}^{x_\mathrm{R}} \, \mathrm{d}x \, e^{\frac{U(x)}{k_B T}}/D(x)}. \label{eq:meanTransitionPathTime}
\end{align}

\subsection{Breakdown of transition-path-time symmetry} 
\begin{figure}
 \centering
 \includegraphics[width=9cm]{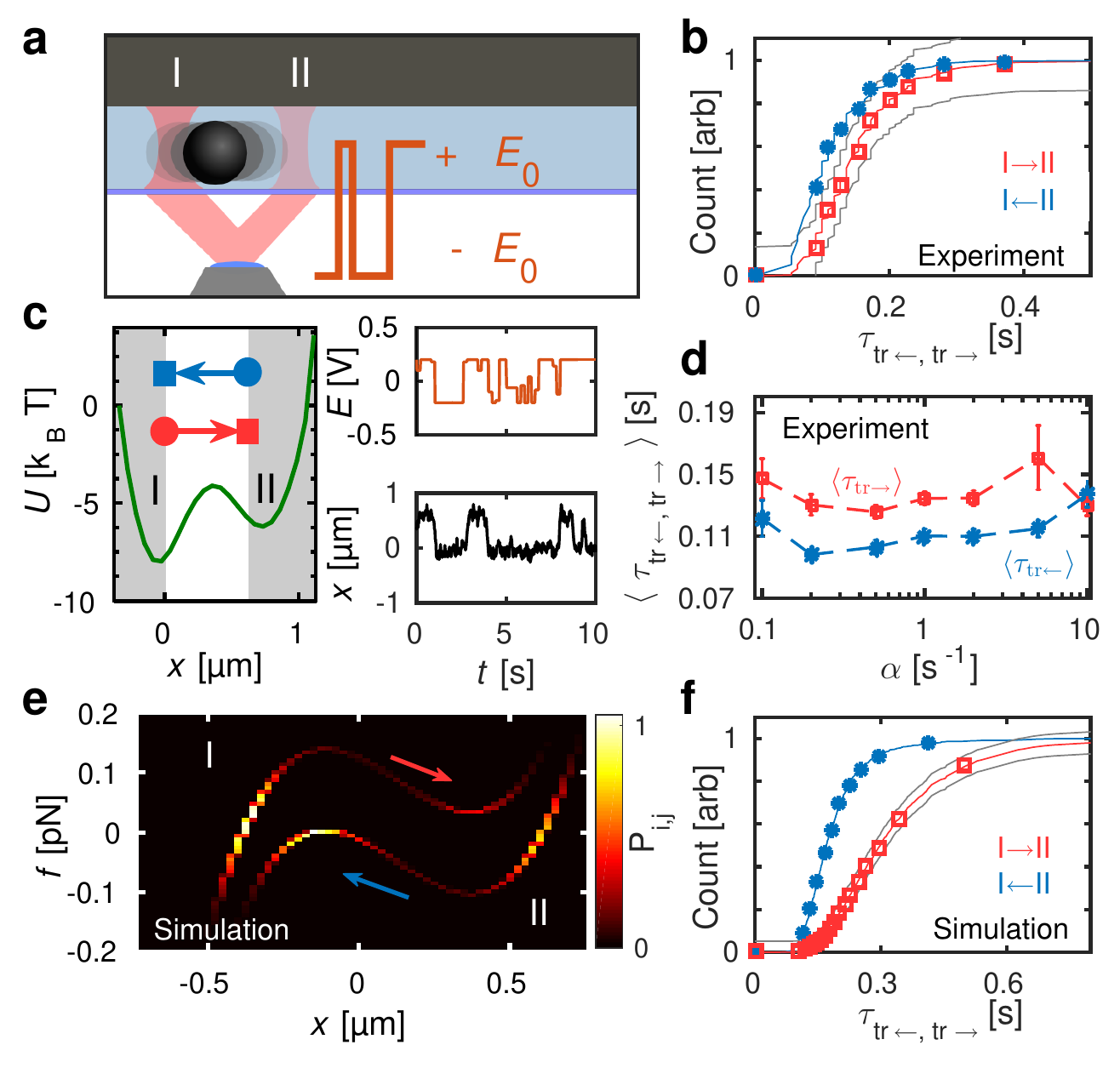}
 \caption{\label{fig:electricalFieldBistable} Transition times of particles between two potential minima in microfluidic experiments and simulations. {\bf a} Bistable potential created by a strong and weak optical trap under stochastically switching voltage $\pm E_0$. {\bf b} Symmetry breaking in the cumulative distribution of transition times $\rm{I}\to \rm{II}$ (red) and $\rm{I} \leftarrow \rm{II}$ (blue). {\bf c} Mean potential $U$ and the first ten seconds of the position $x(t)$ and the measured voltage $E(t)$. Brief jumps in the telegraph process cannot be resolved and appear as averages between the two voltage levels. {\bf d} Mean transition times $\langle \tau_{\leftarrow, \rightarrow}\rangle$ over the decorrelation rate $\alpha$ of the electrical field process. Error bars show the standard error of the mean. {\bf e} Colour map representing the relative occupation of states in the force$\times$position plane under the influence of an external, coloured force. $f$ here denotes the total force $f=f_\text{ext}-\text{d}U/\text{d}x$. {\bf f} Cumulative distribution of transition-path times in Brownian dynamics simulations. }
\end{figure}

The question that arises is whether and how this symmetry can be broken. In the following section, we describe the effect of external forces $f_\text{ext}(t)$, that stochastically switch between two levels $+f_0$ and $-f_0$ with exponentially distributed switching times. This drives the system into a non-equilibrium steady state (NESS). Such two-state switching processes are generally referred to as telegraph noise. The time between two switches is exponentially distributed with a decorrelation rate $\alpha$, such that $\langle f_\text{ext}(t+\Delta t)f_\text{ext}(t) \rangle \propto e^{-\alpha \Delta t}$ for $\Delta t > 0$. We create a NESS on the mesoscale by combining a bistable optical potential consisting of two point traps with different trap strenghts created with our HOT and randomly sign-switching electrical fields (see panel \ref{fig:electricalFieldBistable}{\bf a}). We set the traps apart by $0.7$ \textmu{}m and direct $50$\% more light to the left trap than to the trap on the right, while operating at an overall laser power of $\sim 50$ mW to avoid heat-induced convection. The minima of the two traps correspond to the two states $\rm{I}$ and $\rm{II}$ the transition time is measured over the grey area in panel \ref{fig:electricalFieldBistable}{\bf c}. Different trap strengths result in a difference in curvature between the two traps; the transition barrier loses its symmetry with respect to the centre line separating the two states. 

Indeed, as shown in panel \ref{fig:electricalFieldBistable}{\bf b}, we observe a statistically significant difference in the distribution of transition-path times $\tau_{\rm{I} \to \rm{II}}$ and $\tau_{\rm{I} \leftarrow \rm{II}}$ for $\alpha = 0.5 \, \text{s}^{-1}$. Interestingly, the difference in mean transition times stays fairly constant over a range of decorrelation rates $\alpha$, as shown in panel \ref{fig:electricalFieldBistable}{\bf d}. Towards higher noise decorrelation rates $\alpha$ (note the logarithmic scale), the symmetry is restored. For high frequencies of sign switches, the telegraph force approaches a white noise process and the system should approach the scenario sketched in Fig.~\ref{fig:introSketch}{\bf f}.

To study further the underlying mechanism that led to the breaking of this symmetry, we recreate our experiments in one-dimensional Brownian dynamics simulations. The motion of a colloidal particle under the influence of external telegraph forces is well described by the following equation
\begin{align}
 \gamma \dot{x}(t) &= f_\text{ext}(t) - \frac{\partial U}{\partial x}(x) + \sqrt{2 k_B T \gamma} \xi(t), \label{eq:langevin} 
\end{align}
where $\gamma$ denotes the friction coefficient of a sphere and $f_\text{ext}(t)= f_0\mathcal{T}(t) $ denotes the force exerted by the electrical field. $\mathcal{T}(t)$ represents a random telegraph process that switches between $1$ and $-1$. $U(x)$ corresponds to the free energy and $\xi(t)$ is a Gaussian white-noise process with zero mean and unit variance $\langle \xi(t) \xi(t')\rangle = \delta(t-t')$. We did not attempt to model every parameter of the experiment quantitatively, but rather test the generality of the observed split of transition pathways. The distribution of the system state in the NESS is shown in panel \ref{fig:electricalFieldBistable}{\bf e}. The transition pathways indeed split up. The distribution $P_{i,j}$ of the system state along these pathways in this force$\times$position plane turn out to be visibly different on each leg. Colloids transitioning into one direction will therefore likely experience different force magnitudes along their pathway than colloids transitioning into the opposite direction. 
% This suggests the cause of the broken transition-path time symmetry to be inherently topological. % A colored noise process, such as a telegraph process, can be described in a separate stochastic equation\cite{Gardiner2009}, increasing the dimensionality of the system, such that it can evolve in circular patterns.
The red and blue arrows in the figure indicate the preferred sense of transition direction along the two pathways; detailed balance is indeed broken in this two-dimensional space\cite{Gladrow2016} and transition-path times differ as a consequence (see Fig.~\ref{fig:electricalFieldBistable}{\bf f}).

\subsection{Breakdown of transition-path-time symmetry in DNA-hairpins} Having established broken symmetry on the mesoscale we now demonstrate the generality of the effect with an experimental realisation on the molecular scale. We measure folding and unfolding transition times of short 20-bp DNA-hairpins under the influence of telegraph forces using an optical tweezers-based force spectroscopy setup\cite{Huguet2010} (see Fig.~\ref{fig:DNAHairpin}{\bf a}). The hairpin is grafted onto a colloid of each end. While one colloid is firmly attached to a pipette, the other colloid is held in force-measuring optical tweezers and subject to a feedback-controlled force. If the force is kept constant in time, the hairpin thermally transitions between two main states folded ($F$) and unfolded ($U$), which differ in molecular extension and thus in trap position $\lambda$.
In this paper we describe experiments performed using a  non-equilibrium protocol. Here the system is subject to a telegraph force and each of these two ($U$,$F$) states splits into a doublet: high ($F_+,U_+$) and low force ($F_-,U_-$). This is shown in panel \ref{fig:DNAHairpin}{\bf b} using the density $P_{i,j}$ of states in a coarse-grained space spanned by the force measured by the optical tweezers $f$ and the trap position $\lambda$. Importantly, as the arrows indicate, the telegraph force not only leads to a splitting of states, but also causes transition pathways to diversify. The system is more likely to unfold $F\to U$ during extended periods of high force ($+$), than during periods of low force ($-$). As a consequence, transitions from state $F_-$ to $U_+$ through $U_-$ (red arrows in Fig.~\ref{fig:DNAHairpin}{\bf b},{\bf c}) are more likely than transitions from state $U_+$ to $F_-$ through $F_+$ (blue arrows in Fig.~\ref{fig:DNAHairpin}{\bf b},{\bf c}).

A typical trajectory of the system is shown in panel \ref{fig:DNAHairpin}{\bf c}, highlighting transitions from $U_+$ to $F_-$ (blue) and vice versa (red). This split in pathways results in the visible difference of cumulative distributions of folding (red) and unfolding (blue) transition times in panel \ref{fig:DNAHairpin}{\bf d}. However, this is not necessarily the case: The difference between back and forth transition-path times can become arbitrarily small under certain conditions that we describe in Supplementary Note 4. We conclude that transition-path-time asymmetry points to a lack of information about the system, if not a breakdown of detailed balance. 

Overall, the transition-path times of our DNA-hairpin are significantly longer than previously reported values\cite{Neupane2016}, because our system transitions via intermediate states ($U_-$ or $F_+$). The time spent in corresponding minima affects the overall transition-path time in a path-dependent way and thus amplifies the asymmetry. By contrast, on the mesoscale, transitions are slow enough such that we could resolve the asymmetries shown in Fig.\ref{fig:electricalFieldBistable}{\bf b}, which directly originate from asymmetries in the barrier shape.

We conclude that the overarching topological picture indeed applies to the molecular scale; the dimensionality of the space of folding is effectively increased by one due to the external coloured noise. In this increased phase-space, a breakdown of detailed balance results in a diversification of transition pathways, which causes a transition-path-time asymmetry. Importantly, all participating degrees of freedom, including internal variables of external forces, have to be considered in the analysis.

\section*{Discussion}
\label{sec:discussion}
In our study, we present experimental evidence of a fundamental transition-path-time symmetry in Brownian transitions and its breakdown on the meso- and molecular scale under the influence of stochastic external forces. In accord with intuition, we find that uphill transitions become less likely, as the potential gradient between the initial and end state becomes steeper. Uphill and downhill transition-path times, however, are identically distributed under steady-state conditions. 
Conceptually, we show that transition-path times connecting any two points in the space of the system are thermodynamically well-defined quantities. Indeed, we find that in a time-constant force landscape, measured transition-path times agree with theoretical predictions that assume absorbing boundary conditions at both ends of the transition interval. It is important to note that boundaries can be located anywhere in the potential landscape and do not need to coincide with minima of the potential. 

In contrast to transitions driven by thermal forces, we find that the transition-path-time symmetry can break down under the influence of coloured noise. The additional timescale of external telegraph forces in our systems changes the topology of transition dynamics. We uncover a diversification of transition pathways in the extended phase-space, which includes the external force. Back and forth reactions follow, on average, different paths, breaking detailed balance and the transition-path-time symmetry. Specifically, we show that transition-path times of a colloid in an asymmetric double well potential become measurably asymmetric, when perturbed by randomly switching electrical fields. The asymmetry is sensitive to the frequency of field reversals and disappears for frequencies that are much higher than the barrier crossing time. Similarly, a DNA-hairpin that is driven out of equilibrium by a force that switches randomly between two levels, exhibits asymmetric folding-/unfolding-path times. The observed asymmetry in transition-path times, however, is a result of an implicit projection of the system state onto a one-dimensional reaction coordinate. A breakdown of transition-path-time symmetry does therefore not imply a breakdown of microscopic reversibility. We note that all systems studied here are overdamped and effects related to inertia can be neglected.

Our results have direct implications for the study of transitions in membrane channels or nanopores. Translocation times of solutes, like antibiotics, through membrane channels, are of interest in electrophysiological measurements\cite{Kasianowicz1996, Mahendran2009}. Due to a lack of any direct optical access in these experiments, the shape of the current signal during translocation is the only source of information about the channel-solute interaction. Our work shows that it should be possible to infer the direction of travel solely from first-passage times. A reversal of the electrical potential in such an experiment should result in a distinct translocation time distribution if the solute-channel interaction landscape is asymmetric. The combination of a (sign-flipped) field and interaction potential can be interpreted as the limit of infinite switching times in Fig.~\ref{fig:electricalFieldBistable}, which amounts to the simple case of back and forth translocating solutes experiencing different time-constant force landscapes.  
Furthermore, in studies of molecular motors, transition-path time measurements could enable one to discriminate between power stroke and ratchet mechanisms, beyond thermodynamic considerations\cite{Wang2002}. Arguments based on first-passage-time symmetries have already been used to question the thermodynamic consistency of interpretations of Kinesin motility experiments\cite{Qian2006a, Kolomeisky2005}.

Moreover, in systems driven by ratcheting\cite{Astumian1994, Faucheux1995, Hanggi2009, Skaug2018}, unbiased coloured noise rectifies Brownian dynamics around points of asymmetry of the energy landscape. The asymmetry of transition-path times demonstrated in this study could be used as experimental evidence of this effect. Transition-path-time asymmetries could therefore be helpful in identifying and quantifying non-equilibrium dynamics in biological and molecular systems and complement recently discussed techniques such as broken detailed balance in active matter\cite{Battle2016} and filament fluctuations\cite{Gladrow2016}. We note that a breakdown of the transition-path-time symmetry can be diagnosed by tracking only one degree-of-freedom, whereas diagnosing a breakdown of detailed balance requires a minimal dimension of two in continuous coordinates\cite{Battle2016}. This might be particularly helpful in FRET experiments, where usually only a single degree-of-freedom, the FRET efficiency, is accessible.

The path-time symmetries we explore come in two flavours: an exit-path-time symmetry (see Fig.\ref{fig:microfluidicPanel}) and a transition-path-time symmetry (see Fig.\ref{fig:directTransitionTimes}). Recent theoretical advances\cite{Roldan2015, Neri2016} point to a common origin of both flavours, which lies in a symmetry of the first-passage time of the entropy produced during the transition. Since breakdown of transition-path-time symmetry is a sufficient, but not necessary condition for non-equilibrium dynamics, it is in general not possible to deduce entropy production from observed transition-path times as we discuss in Supplementary Note 4 and Supplementary Figure 4. %The transition-path-time symmetries we report here are directly related to Loschmidt's paradox, which contrasts microscopic reversibility with the macroscopic arrow of time\cite{Berezhkovskii2006}. 

Finally, our results on uphill and downhill first-passage times of Brownian motion are in interesting contrast to a recently reported velocity asymmetry observed in the flow-dominated regime in asymmetrically-shaped microfluidic channels\cite{Siwy2016}. Crucially, the reported asymmetry\cite{Siwy2016} is not an uphill/downhill symmetry in the sense described here, since all observed transitions occurred only into the direction of the applied electrical field. However, a careful investigation of dynamics in the crossover regime of weaker forces would shed light on how such a breakdown of time-reversal symmetry might arise in overdamped driven motion.
%Lastly, in the flow-dominated regime, direction-dependent velocities of colloidal particles have recently been reported in resistive pulse-sensing experiments\cite{Siwy2016}. The symmetry in this study was broken by the asymmetric shape of the micron-sized channels. According to the authors, the origin of the velocity asymmetry lies in a self-focusing effect of particles in the flow as they traverse the channel. Crucially, the reported asymmetry\cite{Siwy2016} is not an uphill/downhill symmetry as investigated in this study, since transitions occurred only in the direction of the applied electrical field.

\section*{Methods}
\label{sec:methods}

\subsection{Microfluidic experiments}
Optical tweezers allow for precise control of Brownian particles on the micro-scale and exert both conservative as well as non-conservative forces. Scattering of photons results in a transfer of photon momentum in the direction of beam propagation and thus gives rise to a mechanical path-dependent force\cite{PhysRevLett.101.128301}. By contrast, forces arising from the gradient of light intensity within the laser beam are path independent and can therefore be interpreted as resulting from an energy potential\cite{AshkinZPressure}. 

In HOTs, computer generated holograms are displayed on spatial light modulators (SLM) to attain almost complete control over the shape of the focal intensity pattern\cite{Curtis2012, grierOpticalTweezerReview}. The technique is highly flexible and permits the creation of several, independent focal shapes at once. Moreover, the phase of the wave front of each individual trap in the focal plane can be addressed such that phase-gradient forces can be applied to particles\cite{Roichman2008}. The resultant forces are non-conservative. In line-shaped traps, the phase-gradient force near the line centre is approximately uniform along the extended direction of the trap. Optical scattering forces lie in the femtoNewton range\cite{Zensen2016} and are thus of ideal magnitude to bias the Brownian motion of colloidal particles over a few micrometers as sketched in Fig.\ref{fig:microfluidicPanel}{\bf a}. Horizontal phase-gradients can be realized in HOTs by laterally shifting the SLM pattern, which imparts a phase-offset onto the beam. The relation between the force exerted by the phase-gradient $f$ and the degree of the shift $p$ turns out to be close to linear for small shifts $|p| \leq 0.4$, which makes the control of forces $f$ significantly easier. The calibration process is described in greater detail in Supplementary Note 1 and Supplementary Figure 1. 

Our SLM-control software has a response time of a few microseconds\cite{Bowman2014}, which permitted us to reliably automate drag-and-drop experiments in a feedback system. The position of all colloids is identified on-line by a peak detection algorithm. The centroid of a small box around the presumed location of the colloid is then calculated from the background subtracted image to refine the position estimate. The optical setup and the microfluidic chip have been described before\cite{Pagliara2013, Pagliara2014}. In contrast to previous publications, we use a Mikrotron MC1362 with adjustable frame rate. The intervals in which measurements are conducted were chosen to be centred in the channel. The microchannels used to confine the colloidal particles all have a length of $4$ \textmu{}m and height of $1$ \textmu{}m. 

The drag-and-drop automation algorithm executes the following steps: (i) The colloid is initialized precisely in the centre of the channel using a HOT point trap. (ii) The point trap holding the colloid is then turned off, a line trap with a prescribed phase-gradient parameter $p$ is turned on and a stopwatch is triggered. (iii) Once the particle leaves a predefined interval, the experiment is stopped and the cycle is repeated with a new phase-gradient parameter. A similar automation procedure was chosen for the two-state experiment with electrical fields that swith polarity randomly, where we varied the rate of switches $\alpha$.

The colloidal particles consist of polycarbonate with a COOH-functionalized surface, with a diameter of $0.5$ \textmu{}m, and these were purchased from Polysciences Inc. The particles were suspended in $0.5\times$TRIS-EDTA buffer at pH 8 and additionally 3 mM KCl to screen potential charges on the walls of the channel.
Prior to each experiment, we ascertained that no hydrodynamic flow was present by comparing left and right exit probabilities of colloidal particles initialized in the centre of the channel.

Electrical fields were created by applying a LabView-controlled voltage signal to silver electrodes that we connected to the far ends of the inlets of our microfluidic chip. 

\subsection{Inference of potentials and diffusion coefficients.}
The forces, $f$, that correspond to each phase-gradient parameter $p$ in Fig.~\ref{fig:microfluidicPanel}{\bf d}-{\bf g} were inferred from the number of left and right exits using the relations $P_\leftarrow$ and $P_\rightarrow$ for exit-path times described in the results section. The excellent agreement of theoretical with experimental exit-path times shows the self-consistency of this approach. We furthermore inferred forces and diffusion coefficients in all microfluidic experiments from Gaussian fits of the form 
\begin{align}
 \rho(\Delta x) = \frac{1}{\sqrt{4\pi D \Delta t}} e^{-\frac{\left (\Delta x - \Delta t f/\gamma \right)^2}{4 D \Delta t} }. \label{eq:gaussInference}
\end{align}
We fitted Eq.~(\ref{eq:gaussInference}) distributions of measured step lengths $\Delta x$, which we parcel into subintervals along the channel. $\Delta t$ in Eq.~(\ref{eq:gaussInference}) is the inverse of the frame rate which was set to $80$ Hz in all microfluidic experiments. We calculated the local friction coefficient $\gamma$ using the Einstein-Stokes equation $D= k_BT/\gamma$. Experiments were performed in a section of the channel where the diffusion coefficient $D$ roughly remains constant\cite{Dettmer2014}, as shown in Supplementary Figure 2{\bf a}.

Due to increased hydrodynamic drag near obstacles, the diffusion coefficient is sensitive to the distance of the colloid to channel walls\cite{Dettmer2014}. Conversely, hydrodynamic drag is reduced close to the entrances of the channel. All one-dimensional diffusion coefficients reported here can be interpreted as averages over y- and z-coordinates. 

The energy potentials $U(x)$ in Figures~\ref{fig:microfluidicPanel}, \ref{fig:directTransitionTimes}, \ref{fig:electricalFieldBistable} were calculated from the force estimates described above by integrating from the left end $x_\mathrm{L}$ of the interval of interest $U(x) = - \int_{x_\mathrm{L}}^x \mathrm{d}x f(x) \approx \sum_i \Delta x/2 (f_{i-1}+f_i)$.

\subsection{Brownian dynamics simulations.}

The Brownian dynamics simulation was set up to qualitatively model the bistable-dynamics experiment. We used a bistable, asymmetric potential of the form $U(x) = a/4x^4+b/2 x^2+cx$, where $c$ controls the asymmetry around $x=0$. The coefficients were set to $a=64 \Delta U_1/L^4$, $b=-a L^2/4$, and $c=2 \Delta U_2/L$, with $\Delta U_1 = 5 \, k_BT $, $\Delta U_2 = 2\, k_BT$ and $L=1$ \textmu{}m. We set the diffusion constant to $D=0.15$ \textmu{}m$^2/s$, which is close to the value a $500$ nm colloid would have in a microchannel (see Supplementary Figure 2). The friction coefficient was obtained, again, using the Einstein-Stokes relation $\gamma = k_BT/D$. The decorrelation time of the telegraph force was set to $2 s$, while the magnitude of the force change was $f_0=\pm 82$ fN.  

\subsection{DNA-hairpin folding and unfolding.}

\begin{figure}
 \centering
 \includegraphics[width=18cm]{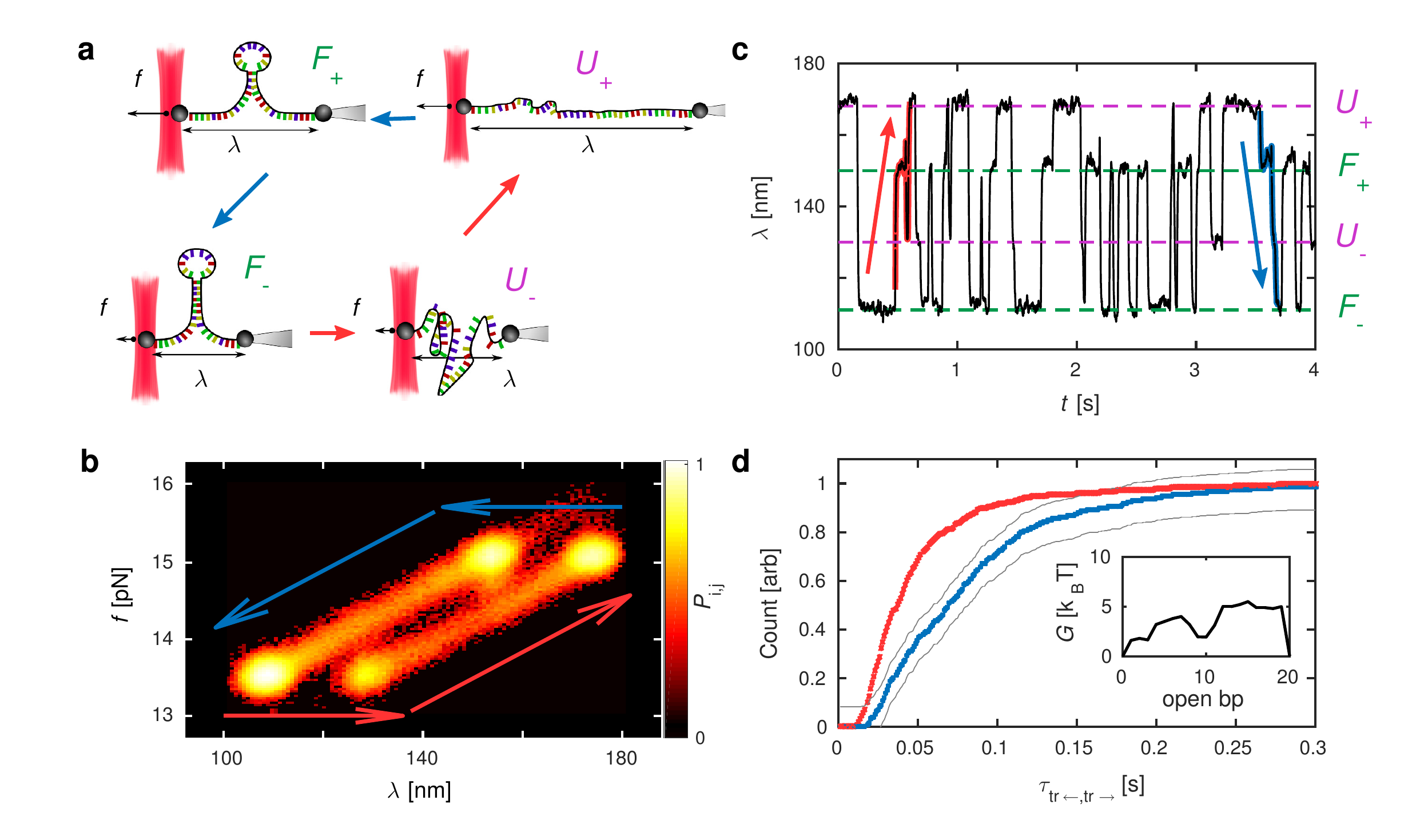}
 \caption{\label{fig:DNAHairpin} Folding and unfolding transitions of DNA hairpins under fluctuating external forces. {\bf a} Sketch of the experiment: Each end of the DNA-hairpin strand is grafted onto a colloid. While one colloid is firmly attached to a pipette, the other is held in force-measuring optical tweezers. The position of the different states $F_+, F_-, U_+, U_-$ are indicated in the $f \times \lambda$ plane (force$\times$extension plane). {\bf b} Relative occupation of states in the $\lambda \times f$ plane. {\bf c} Excerpt of the dynamics of the hairpin. {\bf d} Cumulative histogram of transition times from $U_+ \rightarrow F_-$ (blue) and vice-versa (red). The inset shows the free energy $G$ of the barrier in equilibrium. }
\end{figure}

DNA-hairpin experiments were performed with miniaturized, high-stability optical tweezers equipped with a force-feedback system. The setup uses two counter-propagating focused laser beams ($\lambda = 845\,\text{nm}$, $P = 200\,\text{mW}$) to create a single optical trap. The design of the microfluidic chamber has been described before\cite{Huguet2010}. Force measurements are based on the conservation of linear momentum, and were carried out using Position Sensitive Detectors (PSD) to measure the deflection of the laser beam after interaction with the trapped object\cite{Smith2003}. The position of the trapping beam is monitored by diverting $\sim$ 8\% of each laser beam to a secondary PSD. Altogether the instrument has a resolution of  $0.1 \,\text{pN}$ and $1 \,\text{nm}$ at a 1 kHz acquisition rate. The telegraph forces in Fig.\ref{fig:DNAHairpin} are directly exerted using the optical trap force feedback.

In our experiments, a single DNA hairpin is tethered between two polystyrene beads which are either optically manipulated or trapped by air suction onto the tip of a micropipette. The constructs used in the experiments reported in this paper have short (29 bp) dsDNA handles. The two different handles are differentially labeled with either biotins or digoxigenins. In this way, each handle can selectively bind to either streptavidin ($1.87$ \textmu{}m, Spherotech)  or anti-digoxigenin coated beads ($3.0 - 3.4$ \textmu{}m Kisker Biotech). The synthesis protocols for short (20 bp) DNA hairpins have been previously described\cite{Forns2011}. All experiments were performed at 25$^\circ$C in a buffer containing: 10 mM Tris, 1 mM EDTA, 1 M NaCl, 0.01\% NaN$_3$ (pH 7.5).

\textbf{Data availability} The data used in the current study are available from the corresponding author on reasonable request.

\textbf{Code availability} Custom code used in the current study is available from the corresponding author on reasonable request.

\textbf{Competing interest} The authors declare no competing interests.

\textbf{Author contributions} J.G. wrote the manuscript, carried out HOT experiments, analysed the data and programmed Brownian dynamics simulations. MRC carried out DNA-hairpin experiments, analysed results, contributed ideas and helped write the manuscript. F.R. contributed ideas and designed the DNA-hairpin experiments. U.F.K. designed the HOT experiments and had the initial idea.

\textbf{Acknowledgements} We thank S.M. Bezrukov for stimulating discussions. The research leading to these results has received funding from the European Union’s Horizon 2020 research and innovation program under European Training Network (ETN) Grant No. 674979-NANOTRANS (J.G.). M.R.-C. has received funding from the European Union’s Horizon 2020 research and innovation programme under the Marie Sk\l{}odowska-Curie grant agreement No 749944. J.G. acknowledges the support of the Winton Programme for the Physics of Sustainability. U.F.K. acknowledges funding from an ERC Consolidator Grant (DesignerPores 647144). F.R. acknowledges support from the Spanish Research Council [FIS2016-80458-P]; Catalan Government [Icrea Academia prize 2013], EU [Proseqo, FETOPEN, Proposal 687089].

\bibliography{ref}

\newpage

\end{document}